\documentclass[preprint,amsmath,amssymb,aps,showkeys,showpacs]{revtex4}
\usepackage[english]{babel}
\usepackage{graphicx}
\usepackage{graphics}
\usepackage{amsmath}
\usepackage{dcolumn}
\usepackage{amssymb}
\usepackage{bm}
\usepackage[latin1]{inputenc}

\begin{document}
\title{Relativistic geometric quantum phases from the Lorentz symmetry violation effects in the CPT-even gauge sector of Standard Model Extension}
\author{K. Bakke}
\email{kbakke@fisica.ufpb.br}
\affiliation{Departamento de F\'isica, Universidade Federal da Para\'iba, Caixa Postal 5008, 58051-970, Jo\~ao Pessoa, PB, Brazil.}

\author{H. Belich} 
\email{belichjr@gmail.com}
\affiliation{Departamento de F\'isica e Qu\'imica, Universidade Federal do Esp\'irito Santo, Av. Fernando Ferrari, 514, Goiabeiras, 29060-900, Vit\'oria, ES, Brazil.}

\begin{abstract}
We discuss the appearance of relativistic geometric quantum phases for a Dirac neutral particle based on possible scenarios of the Lorentz symmetry violation background in the CPT-even gauge sector of Standard Model Extension. Relativistic analogues of the Anandan quantum phase [J. Anandan, Phys. Lett. A {\bf138}, 347 (1989)] are obtained based on the the parity-even and parity-odd sectors of the tensor $\left(K_{F}\right)_{\mu\nu\alpha\beta}$. 
\end{abstract}

\keywords{Lorentz symmetry violation, geometric phase, Dirac neutral particles, Anandan quantum phase, Aharonov-Casher effect, He-McKellar-Wilkens effect, scalar Aharonov-Bohm effect}
\pacs{11.30.Cp, 03.65.Pm, 03.65.Ge, 03.65.Vf}

\maketitle

\section{Introduction}

An interesting point of discussion in the area of particle physics is the origin of electron electric dipole moment which has not been explained by Standard Model of particle physics yet. At present days, it is well-known that just experimental upper bounds have been established \cite{revmod}. Based on the Standard Model, an upper limit for the electric dipole moment of an electron was established as $d_{e}\leq10^{-38}\,\mathrm{e}\cdot\mathrm{cm}$ \cite{revmod}. On the other hand, experiments measured an upper limit given by $d_{e}\leq10^{-29}\,\mathrm{e}\cdot\mathrm{cm}$ by using a polar molecule thorium monoxide (ThO) \cite{science}. This experimental result has shown a necessity of investigating the physics beyond the Standard Model because the term associated with the electric dipole moment violates the CP symmetry. A possible way of dealing with a scenario beyond the Standard Model is the extension of the mechanism for spontaneous symmetry breaking through vector or tensor fields, which implies that the Lorentz symmetry is violated.

Models that deal with the violation of the Lorentz symmetry have appeared in the literature after the seminal work made by Kosteleck\'{y} and Samuel \cite{extra3} in the string theory, where it is shown that the Lorentz symmetry is violated through a spontaneous symmetry breaking mechanism triggered by the appearance of nonvanishing vacuum expectation values of nontrivial Lorentz tensors. Such models are considered as effective theories whose analysis of the phenomenological aspect at low energies may provide information and impose restrictions on the fundamental theory in which they stem from. Progresses of a possible generalization of the Standard Model that includes the spontaneous violation of the Lorentz symmetry have allowed us to build a general framework for testing the low-energy manifestations of CPT symmetry and the Lorentz symmetry breaking which is known as the Standard Model Extension (SME) \cite{colladay-kost}. In this framework, the effective Lagrangian corresponds to the usual Lagrangian of the Standard Model to which is added to the Standard Model operators a Lorentz violation tensor background. The effective Lagrangian is written as an invariant under the Lorentz transformation of coordinates in order to guarantee that the observer independence of physics. However, the physically relevant transformations are those that affect only the dynamical fields of the theory. These changes are called particle transformations, whereas the coordinate transformations (including the background tensor) are called the observer transformations. In Refs. \cite{coll-kost,baeta,bras}, one can find a deep analysis of these concepts. Concerning the experimental searches for the CPT/Lorentz-violation signals, the generality of the SME has provided the basis for many investigations. In the flat spacetime limit, empirical studies include muons \cite{muon}, mesons \cite{meson,meson2}, baryons \cite{barion}, photons \cite{photon}, electrons \cite{electron}, neutrinos \cite{neutrino} and the Higgs sector \cite{higgs}. The gravity sector has also been explored in Refs. \cite{gravity,gravity2}. In Ref. \cite{data}, one can find the current limits on the coefficients of the Lorentz symmetry violation. In recent years, Lorentz symmetry breaking effects have been investigated in the hydrogen atom \cite{manoel}, on the Rashba coupling \cite{rash,bb3}, in a quantum ring \cite{bb4}, in Weyl semi-metals \cite{weyl}, in tensor backgrounds \cite{louzada,manoel2}, in the quantum Hall effect \cite{lin2} and geometric quantum phases \cite{belich,belich1,bb2,bbs3,lbb,bbs2}.

The arising of geometric quantum phases in interferometry experiments stems from the presence of a potential vector along the path of a charged particle even though there exists no interaction with a magnetic field \cite{ab} or electric field \cite{dab}. Geometric phases was introduced by Berry \cite{berry} in 1984 to describe the phase shift acquired by the wave function of a quantum particle in an adiabatic cyclic evolution. At present days, it is well-known that geometric quantum phases can be measured in any cyclic evolution \cite{ahan}. The best famous quantum effect related to the appearance of geometric phases is the Aharonov-Bohm effect (AB) \cite{ab}. It is worth mentioning other quantum effects related to geometric phases that are termed the dual effect of the Aharonov-Bohm effect \cite{dab} and the scalar Aharonov-Bohm effect \cite{pesk}. 

In recent years, the study of geometric quantum phases has been extended to neutral particles with permanent magnetic dipole moment \cite{ac} and permanent electric dipole moment \cite{hmw}. However, the quantum effects associated with geometric phases for neutral particles stem from the interaction between the magnetic (electric) dipole moment of the neutral particle with electric (magnetic) field and are considered an AB-type effect in the sense that this interaction is a force-free interaction \cite{disp3}. Well-known quantum effects associated with geometric phases for neutral particle are the Aharonov-Casher effect \cite{ac}, the He-McKellar-Wilkens effect \cite{hmw} and the scalar AB effect for neutral particles \cite{zei,anan,anan2}. Recently, analogues effects for neutral particle have been studied, such as analogues of the He-McKellar-Wilkens effect \cite{furt,bbs3}, analogues of the Aharonov-Casher effect \cite{furt2,lbb} and analogues of the scalar Aharonov-Bohm effect for neutral particles \cite{anan,anan2,zei,bbs3}.

In this paper, we study the appearance of relativistic geometric quantum phases in the wave function of a Dirac neutral particle based on possible scenarios of the Lorentz symmetry breaking defined by a tensor $\left(K_{F}\right)_{\mu\nu\alpha\beta}$ that corresponds to a tensor that governs the Lorentz symmetry violation in the CPT-even gauge sector of the Standard Model Extension. We show that relativistic analogues of the Anandan quantum phase \cite{anan,anan2} can be obtained and, as particular cases, relativistic analogues of the Aharonov-Casher effect \cite{ac}, the He-McKellar-Wilkens effect \cite{hmw} and the scalar Aharonov-Bohm effect for neutral particles \cite{zei,anan,anan2} can be achieved. Further, we estimate upper bounds for constant parameters of the Lorentz symmetry violation in particular cases.

This paper is organized as follows: in section II, we introduce the Lorentz symmetry violation background defined by a tensor that governs the Lorentz symmetry violation in the CPT-even gauge sector of the Standard Model Extension; in section III, we discuss the appearance of relativistic geometric phases induced by the parity-even sector of the tensor $\left(K_{F}\right)_{\mu\nu\alpha\beta}$ that governs the Lorentz symmetry violation; in section IV, we discuss the appearance of relativistic geometric phases induced by the parity-odd sector of the tensor $\left(K_{F}\right)_{\mu\nu\alpha\beta}$ that governs the Lorentz symmetry violation; in section V, we present our conclusions.

\section{Background of the Lorentz symmetry violation}

In recent years, a nonminimal coupling has been suggested in the fermionic sector in order to study the behaviour of fermions beyond the description of the Standard Model. This nonminimal coupling has an interesting property where the effects of the Lorentz symmetry breaking in the CPT-even Gauge Sector of the Standard Model Extension is governed by by a tensor background. Such coupling allows us to write the Dirac equation in the form \cite{louzada,manoel2}: 
\begin{eqnarray}
i\gamma^{\mu}\partial_{\mu}\psi+\frac{ig}{2}\gamma^{\mu}\left(K_{F}\right)_{\mu\nu\alpha\beta}\,\gamma^{\nu}\,F^{\alpha\beta}\left(x\right)\,\psi=m\psi,
\label{1}
\end{eqnarray}
where $g$ is a constant, $\left(K_{F}\right)_{\mu\nu\alpha\beta}$ corresponds to a tensor that governs the Lorentz symmetry violation in the CPT-even gauge sector of the Standard Model Extension. In particular, the tensor $\left(K_{F}\right)_{\mu\nu\alpha\beta}$ can be written in terms of four $3\times3$ matrices defined as
\begin{eqnarray}
\left(\kappa_{DE}\right)_{ij}&=&-2\left(K_{F}\right)_{0j0k};\nonumber\\
\left(\kappa_{HB}\right)_{jk}&=&\frac{1}{2}\epsilon_{jpq}\,\epsilon_{klm}\left(K_{F}\right)^{pqlm}\\
\left(\kappa_{DB}\right)_{jk}&=&-\left(\kappa_{HE}\right)_{kj}=\epsilon_{kpq}\left(K_{F}\right)^{0jpq}.\nonumber
\label{eq:}
\end{eqnarray}
Observe that the matrices $\left(\kappa_{DE}\right)_{ij}$ and $\left(\kappa_{HB}\right)_{ij}$ are symmetric and represent the parity-even sector of the tensor $\left(K_{F}\right)_{\mu\nu\alpha\beta}$. On the other hand, the matrices $\left(\kappa_{DB}\right)_{ij}$ and $\left(\kappa_{HE}\right)_{ij}$ has no symmetry and represent the parity-odd sector of the tensor $\left(K_{F}\right)_{\mu\nu\alpha\beta}$.

Furthermore, the tensor $F_{\mu\nu}\left(x\right)$ in (\ref{1}) corresponds to the usual electromagnetic tensor ($F_{0i}=-F_{i0}=E_{i}$, and $F_{ij}=-F_{ji}=\epsilon_{ijk}B^{k}$), and the $\gamma^{\mu}$ matrices are defined in the Minkowski spacetime in the form \cite{greiner}:
\begin{eqnarray}
\gamma^{0}=\hat{\beta}=\left(
\begin{array}{cc}
1 & 0 \\
0 & -1 \\
\end{array}\right);\,\,\,\,\,\,
\gamma^{i}=\hat{\beta}\,\hat{\alpha}^{i}=\left(
\begin{array}{cc}
 0 & \sigma^{i} \\
-\sigma^{i} & 0 \\
\end{array}\right);\,\,\,\,\,\,\Sigma^{i}=\left(
\begin{array}{cc}
\sigma^{i} & 0 \\
0 & \sigma^{i} \\	
\end{array}\right),
\label{2}
\end{eqnarray}
with $\vec{\Sigma}$ being the spin vector. The matrices $\sigma^{i}$ correspond to the Pauli matrices and satisfy the relation $\left(\sigma^{i}\,\sigma^{j}+\sigma^{j}\,\sigma^{i}\right)=2\eta^{ij}$. Our interest is to work with curvilinear coordinates, thus, we need to apply a coordinate transformation $\frac{\partial}{\partial x^{\mu}}=\frac{\partial \bar{x}^{\nu}}{\partial x^{\mu}}\,\frac{\partial}{\partial\bar{x}^{\nu}}$, and a unitary transformation on the wave function $\psi\left(x\right)=U\,\psi'\left(\bar{x}\right)$ \cite{schu,bbs3}. In this way, the Dirac equation can be written in any orthogonal system in the presence of Lorentz symmetry breaking effects described in Eq. (\ref{1}) as \cite{bbs3} 
\begin{eqnarray}
i\,\gamma^{\mu}\,D_{\mu}\,\psi+\frac{i}{2}\,\sum_{k=1}^{3}\,\gamma^{k}\,\left[D_{k}\,\ln\left(\frac{h_{1}\,h_{2}\,h_{3}}{h_{k}}\right)\right]\psi+\frac{ig}{2}\gamma^{\mu}\left(K_{F}\right)_{\mu\nu\alpha\beta}\,\gamma^{\nu}\,F^{\alpha\beta}\left(x\right)\psi=m\psi,
\label{4}
\end{eqnarray}
where $D_{\mu}=\frac{1}{h_{\mu}}\,\partial_{\mu}$ is the derivative of the corresponding coordinate system, and the parameters $h_{k}$ correspond to the scale factors of this coordinate system \cite{schu}. For instance, the line element of the Minkowski spacetime is writing in cylindrical coordinates in the form:
\begin{eqnarray}
ds^{2}=-dt^{2}+d\rho^{2}+\rho^{2}d\varphi^{2}+dz^{2},
\label{eq:}
\end{eqnarray}
then, the corresponding scale factors are $h_{0}=1$, $h_{1}=1$, $h_{2}=\rho$ and $h_{3}=1$ \cite{schu}. Moreover, the second term in (\ref{4}) gives rise to a term called the spinorial connection $\Gamma_{\mu}\left(x\right)$ \cite{schu,bb2,bbs3,bbs2}. In this way, the Dirac equation (\ref{4}) becomes \cite{bbs3}
\begin{eqnarray}
m\psi&=&i\gamma^{0}\frac{\partial\psi}{\partial t}+i\gamma^{1}\left(\frac{\partial}{\partial\rho}+\frac{1}{2\rho}\right)\psi+i\frac{\gamma^{2}}{\rho}\,\frac{\partial\psi}{\partial\varphi}+i\gamma^{3}\frac{\partial\psi}{\partial z}\nonumber\\
[-2mm]\label{5}\\[-2mm]
&+&ig\,\vec{\alpha}\cdot\vec{\mathbb{E}}\,\psi-g\,\vec{\Sigma}\cdot\vec{\mathbb{B}}\,\psi+ig\,\vec{\alpha}\cdot\vec{\mathbf{B}}\,\psi-g\,\vec{\Sigma}\cdot\vec{\mathbf{E}}\,\psi,\nonumber
\end{eqnarray}
where we have defined effective fields in the Dirac equation (\ref{5}) as \cite{lbb}
\begin{eqnarray}
\mathbb{E}_{i}&=&\left(\kappa_{DE}\right)_{ij}\,E^{j};\,\,\,\,\,\,\mathbb{B}_{i}=\left(\kappa_{HB}\right)_{ij}\,B^{j};\nonumber\\
[-2mm]\label{6}\\[-2mm]
\mathbf{E}_{i}&=&\left(\kappa_{HE}\right)_{ij}\,E^{j};\,\,\,\,\,\,\mathbf{B}_{i}=\left(\kappa_{DB}\right)_{ij}\,B^{j}.\nonumber
\end{eqnarray}  

Hence, from the definition of the effective fields given in Eq. (\ref{6}), we show that different scenarios of the Lorentz symmetry violation can be built in order to study the appearance of relativistic geometric phases in the wave function of a Dirac neutral particle. Our objective in the next sections is to obtain relativistic analogues of the Anandan quantum phase \cite{anan,anan2}, the Aharonov-Casher effect \cite{ac}, the He-McKellar-Wilkens effect \cite{hmw} and the scalar Aharonov-Bohm effect for neutral particles \cite{zei,anan,anan2}.

\section{relativistic geometric phases induced by the parity-even sector of the tensor $\left(K_{F}\right)_{\mu\nu\alpha\beta}$}

In this section, our interest is to obtain relativistic geometric quantum phases that appears in the wave function of a Dirac neutral particle induced by the parity-even sector of the tensor $\left(K_{F}\right)_{\mu\nu\alpha\beta}$ that governs the Lorentz symmetry violation. We start by considering only the matrices that represent the parity-even sector of the tensor $\left(K_{F}\right)_{\mu\nu\alpha\beta}$ can have non-null components. In this way, the Dirac equation (\ref{5}) becomes
\begin{eqnarray}
m\psi=i\gamma^{0}\frac{\partial\psi}{\partial t}+i\gamma^{1}\left(\frac{\partial}{\partial\rho}+\frac{1}{2\rho}\right)\psi+i\frac{\gamma^{2}}{\rho}\,\frac{\partial\psi}{\partial\varphi}+i\gamma^{3}\frac{\partial\psi}{\partial z}+ig\,\vec{\alpha}\cdot\vec{\mathbb{E}}\,\psi-g\,\vec{\Sigma}\cdot\vec{\mathbb{B}}\,\psi.
\label{3.1}
\end{eqnarray}

By applying the Dirac phase factor method \cite{dirac,dirac2} into the Dirac equation (\ref{3.1}), we can write the Dirac spinor in the form $\psi=e^{i\phi}\,\psi_{0}$, where $\psi_{0}$ is the solution of the equation
\begin{eqnarray}
m\psi_{0}=i\gamma^{0}\frac{\partial\psi_{0}}{\partial t}+i\gamma^{1}\left(\frac{\partial}{\partial\rho}+\frac{1}{2\rho}\right)\psi_{0}+i\frac{\gamma^{2}}{\rho}\,\frac{\partial\psi_{0}}{\partial\varphi}+i\gamma^{3}\frac{\partial\psi_{0}}{\partial z},
\label{3.2}
\end{eqnarray} 
and the general expression for the geometric phase acquired by the wave function of the neutral particle is given by
\begin{eqnarray}
\phi_{\mathrm{A}}=-g\,\hat{\beta}\,\oint\left[\vec{\Sigma}\times\vec{\mathbb{E}}\right]\cdot d\vec{r}-g\,\hat{\beta}\,\int_{0}^{\tau}\,\vec{\Sigma}\cdot\vec{\mathbb{B}}\,dt.
\label{3.3}
\end{eqnarray}

The relativistic geometric phase given in (\ref{3.3}) corresponds to the relativistic analogue of the Anandan geometric phase for a neutral particle with a permanent magnetic dipole moment \cite{anan,anan2} based on a Lorentz symmetry breaking scenario defined by the parity-even sector of the tensor $\left(K_{F}\right)_{\mu\nu\alpha\beta}$. In contrast to a previous study of the relativistic Anandan quantum phase induced by a Lorentz symmetry violation background \cite{bbs2}, whose geometric phase is an Abelian geometric phase, we have that the relativistic Anandan quantum phase (\ref{3.3}) is a non-Abelian geometric phase. This difference between the Abelian nature and the non-Abelian nature of the relativistic geometric phases obtained in Ref. \cite{bbs2} and in Eq. (\ref{3.3}) results from the Lorentz symmetry violation background which is defined by a tensor field in (\ref{1}). 

Let us consider a particular case defined by a field configuration given by a radial electric field produced by a uniform linear distribution of electric charges on the $z$-axis, that is, $\vec{E}=E^{1}\,\hat{\rho}=\frac{\lambda}{\rho}\,\hat{\rho}$, where $\lambda$ corresponds to the linear density of electric charges, $\rho=\sqrt{x^{2}+y^{2}}$ is the radial coordinate, and $\hat{\rho}$ is a unit vector in the radial direction. In this case, the relativistic Anandan quantum phase (\ref{3.3}) becomes
\begin{eqnarray}
\phi_{1}=-\zeta_{1}\,\hat{\beta}\,\Sigma^{3}+\zeta_{2}\,\hat{\beta}\,\Sigma^{1}.
\label{3.4}
\end{eqnarray} 
where $\zeta_{1}=2\pi\lambda\,g\left(\kappa_{DE}\right)_{11}$ and $\zeta_{2}=2\pi\lambda\,g\left(\kappa_{DE}\right)_{31}$. An interesting case occurs if we consider the matrix $\left(\kappa_{DE}\right)_{ij}$ to be a diagonal matrix. From this, we have
\begin{eqnarray}
\phi_{\mathrm{AC}}=-2\pi\lambda\,g\left(\kappa_{DE}\right)_{11}\,\hat{\beta}\,\Sigma^{3},
\label{3.5}
\end{eqnarray} 
which corresponds to the relativistic analogue of the Aharonov-Casher effect \cite{ac} based on a Lorentz symmetry breaking scenario defined by the parity-even sector of the tensor $\left(K_{F}\right)_{\mu\nu\alpha\beta}$. Observe that the relativistic geometric phase (\ref{3.5}) does not depend on the velocity of the Dirac neutral particle which consists in a non-dispersive geometric phase as established in Refs. \cite{disp,disp2,disp3}. Note that the relativistic analogue of the Aharonov-Casher effect geometric phase given in (\ref{3.5}) is a non-Abelian geometric phase due to the Lorentz symmetry violation background defined by a tensor field in (\ref{1}) in contrast to the results of Ref. \cite{bbs2}, whose analogue of the Aharonov-Casher geometric phase \cite{ac} is an Abelian geometric phase because the Lorentz symmetry violation background is defined by a fixed vector field.  

Let us suppose an experimental ability to measure geometrical phases as small as $10^{-4}\,\mathrm{rad}$ \cite{cimino}, then, we can affirm that the theoretical phase induced for a neutral particle cannot be larger than this value, that is, $\left|\phi_{\mathrm{AC}}\right|<10^{-4}\mathrm{rad}$. Thereby, by considering the spin of the neutral particle to be up, by taking $\left|\vec{E}\right|\approx10^{7}\,\mathrm{V}/\mathrm{m}$ and $r_{0}=10^{-5}\,\,\mathrm{m}$ (which correspond to the usual values of electric fields and radius for 1D mesoscopic rings \cite{Nitta}), we can estimate an upper bound for the constant $g\left(\kappa_{DE}\right)_{11}$ given by
\begin{eqnarray}
\left|g\left(\kappa_{DE}\right)_{11}\right|\,<\,2.2\times10^{-9}\,\left(\mathrm{eV}\right)^{-1}.
\label{3.6}
\end{eqnarray}

Next, let us consider another particular case defined by a field configuration given by a uniform magnetic field on the $z$-axis, $\vec{B}=B_{0}\,\hat{z}$. In this case, the relativistic Anandan quantum phase (\ref{3.3}) becomes
\begin{eqnarray}
\phi_{2}=q_{1}\,\hat{\beta}\,\Sigma^{3}-q_{2}\,\hat{\beta}\,\Sigma^{2}-q_{3}\,\hat{\beta}\,\Sigma^{1},
\label{3.7}
\end{eqnarray}
where we have defined the parameters $q_{1}=g\,\left(\kappa_{HB}\right)_{33}\,B_{0}\tau$, $q_{2}=g\,\left(\kappa_{HB}\right)_{23}\,B_{0}\tau$ and $q_{3}=g\,\left(\kappa_{HB}\right)_{13}\,B_{0}\tau$. It is interesting to note that if we consider the matrix $\left(\kappa_{HB}\right)_{ij}$ to be a diagonal matrix, the relativistic geometric phase (\ref{3.7}) becomes
\begin{eqnarray}
\phi_{\mathrm{SAB}}=-g\,\left(\kappa_{HB}\right)_{33}\,B_{0}\tau\,\hat{\beta}\,\Sigma^{3}.
\label{3.8}
\end{eqnarray}

The quantum effect associated to the phase shift given in (\ref{3.8}) corresponds to the relativistic analogue of the scalar Aharonov-Bohm effect for a neutral particle with a permanent magnetic dipole moment \cite{zei,anan,anan2} based on a Lorentz symmetry breaking scenario defined by the parity-even sector of the tensor $\left(K_{F}\right)_{\mu\nu\alpha\beta}$. Let us suppose again an experimental ability to measure geometrical phases as small as $10^{-4}\,\mathrm{rad}$ \cite{cimino}, with $B_{0}=21\times10^{-4}\,\mathrm{T}$ and $\tau=17,8\times10^{-6}\,\mathrm{s}$ \cite{sab}. From this, we can estimate an upper bound for the constant $g\left(\kappa_{HB}\right)_{33}$ given by
\begin{eqnarray}
\left|g\left(\kappa_{HB}\right)_{33}\right|\,<\,3.6\times 10^{-32}\,\left(\mathrm{eV}\right)^{-1}.
\label{3.9}
\end{eqnarray}

Finally, we must observe that the relativistic geometric phase (\ref{3.8}) is a non-Abelian geometric phase and it is a non-dispersive phase because it does not depend on the velocity of the Dirac neutral particle \cite{disp,disp2,disp3}.

Hence, we have seen that relativistic analogues of the Anandan quantum phase \cite{anan,anan2}, the Aharonov-Casher effect \cite{ac} and the scalar Aharonov-Bohm effect for neutral particles with a permanent magnetic dipole moment  \cite{zei,anan,anan2} can be obtained by defining different scenarios of the Lorentz symmetry breaking induced by the parity-even sector of the tensor $\left(K_{F}\right)_{\mu\nu\alpha\beta}$. Moreover, we have seen that upper bounds for constant parameters of the Lorentz symmetry breaking can be estimated based on these relativistic analogue effects.

\section{relativistic geometric phase induced by the parity-odd sector of the tensor $\left(K_{F}\right)_{\mu\nu\alpha\beta}$}

In this section, we consider only the matrices that represent the parity-odd sector of the tensor $\left(K_{F}\right)_{\mu\nu\alpha\beta}$ can have non-null components. The main objective is to obtain relativistic geometric quantum phases induced by the parity-odd sector of the tensor $\left(K_{F}\right)_{\mu\nu\alpha\beta}$ that governs the Lorentz symmetry violation. Thereby, the Dirac equation (\ref{5}) becomes
\begin{eqnarray}
m\psi=i\gamma^{0}\frac{\partial\psi}{\partial t}+i\gamma^{1}\left(\frac{\partial}{\partial\rho}+\frac{1}{2\rho}\right)\psi+i\frac{\gamma^{2}}{\rho}\,\frac{\partial\psi}{\partial\varphi}+i\gamma^{3}\frac{\partial\psi}{\partial z}+ig\,\vec{\alpha}\cdot\vec{\mathbf{B}}\,\psi-g\,\vec{\Sigma}\cdot\vec{\mathbf{E}}\,\psi.
\label{4.1}
\end{eqnarray}

By applying the Dirac phase factor method \cite{dirac,dirac2} into the Dirac equation (\ref{4.1}), we can write the Dirac spinor in the form $\psi=e^{i\phi}\,\psi_{0}$, then, we have that $\psi_{0}$ is the solution of the equation
\begin{eqnarray}
m\psi_{0}=i\gamma^{0}\frac{\partial\psi_{0}}{\partial t}+i\gamma^{1}\left(\frac{\partial}{\partial\rho}+\frac{1}{2\rho}\right)\psi_{0}+i\frac{\gamma^{2}}{\rho}\,\frac{\partial\psi_{0}}{\partial\varphi}+i\gamma^{3}\frac{\partial\psi_{0}}{\partial z},
\label{4.2}
\end{eqnarray} 
and the general expression for the geometric phase acquired by the wave function of the neutral particle is given by
\begin{eqnarray}
\phi_{\mathrm{A}}'=-g\,\hat{\beta}\,\oint\left[\vec{\Sigma}\times\vec{\mathbf{B}}\right]\cdot d\vec{r}-g\,\hat{\beta}\,\int_{0}^{\tau}\,\vec{\Sigma}\cdot\vec{\mathbf{E}}\,dt.
\label{4.3}
\end{eqnarray}

The relativistic geometric phase given in (\ref{4.3}) corresponds to the relativistic analogue of the Anandan geometric phase for a neutral particle with a permanent electric dipole moment \cite{anan,anan2} based on a Lorentz symmetry breaking scenario defined by the parity-odd sector of the tensor $\left(K_{F}\right)_{\mu\nu\alpha\beta}$. Again, we can see that the relativistic Anandan quantum phase is a non-Abelian geometric phase in contrast to the results of Ref. \cite{bbs3}, whose Anandan quantum phase is an Abelian geometric phase. As we have discussed in the previous section, this difference between the non-Abelian nature and the Abelian nature also results from the Lorentz symmetry violation background is defined by a tensor field background given in Eq. (\ref{1}).

Let us consider a special case defined by a field configuration given by a radial magnetic field produced by a uniform linear distribution of magnetic charges on the $z$-axis, that is, $\vec{B}=B^{1}\,\hat{\rho}=\frac{\lambda_{m}}{\rho}\,\hat{\rho}$, where $\lambda_{m}$ corresponds to the linear density of magnetic charges, $\rho=\sqrt{x^{2}+y^{2}}$ is the radial coordinate, and $\hat{\rho}$ is a unit vector in the radial direction. The experimental possible apparatus for achieving this field configuration was discussed in Ref. \cite{tka}. In this case, the relativistic Anandan quantum phase (\ref{4.3}) becomes
\begin{eqnarray}
\phi_{3}=-a_{1}\,\hat{\beta}\,\Sigma^{3}+a_{2}\,\hat{\beta}\,\Sigma^{1},
\label{4.4}
\end{eqnarray} 
where $a_{1}=2\pi\lambda_{m}\,g\left(\kappa_{DB}\right)_{11}$ and $a_{2}=2\pi\lambda\,g\left(\kappa_{DB}\right)_{31}$. Observe that if we consider the matrix $\left(\kappa_{DB}\right)_{ij}$ to be a diagonal matrix, we have
\begin{eqnarray}
\phi_{\mathrm{HMW}}=-2\pi\lambda_{m}\,g\left(\kappa_{DB}\right)_{11}\,\hat{\beta}\,\Sigma^{3},
\label{4.5}
\end{eqnarray} 
which corresponds to the relativistic analogue of the He-McKellar-Wilkens effect \cite{hmw} based on a Lorentz symmetry breaking scenario defined by the parity-odd sector of the tensor $\left(K_{F}\right)_{\mu\nu\alpha\beta}$. It is hard to estimate upper bounds for parameters of the Lorentz symmetry breaking based on the relativistic analogue of the He-McKellar-Wilkens effect given in (\ref{4.5}), since magnetic charge have not been observed. However, experiments have been proposed in order to reproduce the He-McKellar-Wilkens field configuration \cite{tka,tka1,tka2,tka3}. Recently, an experimental test for the He-McKellar-Wilkens geometric phase has been made in atom interferometry \cite{hmw2}. Despite the experiments cited above can reproduce the He-McKellar-Wilkens field configuration, we decided do not discuss any upper bound for the Lorentz symmetry breaking parameters in order to avoid any misunderstanding with our assumption of magnetic charges.

Further, we can also observe that the relativistic analogue of the He-McKellar-Wilkens effect \cite{hmw} given in Eq. (\ref{4.5}) is a non-dispersive geometric phase as established in Refs. \cite{disp,disp2,disp3}. In addtiion, the geometric phase (\ref{4.5}) is a non-Abelian geometric phase due to the Lorentz symmetry violation background defined by a tensor field in Eq. (\ref{1}). From the Lorentz symmetry violation background defined by a fixed vector field, by contrast, the analogue of the He-McKellar-Wilkens effect \cite{hmw} obtained in Ref. \cite{bbs3} is an Abelian and a non-dispersive geometric phase.

Finally, let us consider another special case defined by a field configuration given by a uniform electric field on the $z$-axis, $\vec{E}=E_{0}\,\hat{z}$. In this case, the relativistic Anandan quantum phase (\ref{4.3}) becomes
\begin{eqnarray}
\phi_{4}=d_{1}\,\hat{\beta}\,\Sigma^{3}-d_{2}\,\hat{\beta}\,\Sigma^{2}-d_{3}\,\hat{\beta}\,\Sigma^{1},
\label{4.8}
\end{eqnarray}
where we have defined the parameters $d_{1}=g\,\left(\kappa_{HE}\right)_{33}\,E_{0}\tau$, $d_{2}=g\,\left(\kappa_{HE}\right)_{23}\,E_{0}\tau$ and $d_{3}=g\,\left(\kappa_{HE}\right)_{13}\,E_{0}\tau$. It is interesting to note that if we consider the matrix $\left(\kappa_{HE}\right)_{ij}$ to be a diagonal matrix, the relativistic geometric phase (\ref{4.8}) becomes
\begin{eqnarray}
\phi_{\mathrm{SAB}}'=-g\,\left(\kappa_{HE}\right)_{33}\,E_{0}\tau\,\hat{\beta}\,\Sigma^{3}.
\label{4.9}
\end{eqnarray}

The quantum effect associated to the phase shift given in (\ref{4.9}) corresponds to the relativistic analogue of the scalar Aharonov-Bohm effect for a neutral particle with a permanent electric dipole moment \cite{zei,anan,anan2} based on a Lorentz symmetry breaking scenario defined by the parity-odd sector of the tensor $\left(K_{F}\right)_{\mu\nu\alpha\beta}$.  Observe that the relativistic geometric phase (\ref{4.9}) is also a non-Abelian geometric phase. The geometric phase (\ref{4.9}) is also a non-dispersive phase because it does not depend on the velocity of the Dirac neutral particle as discussed in Refs. \cite{disp,disp2,disp3}.

Hence, we have shown that relativistic analogues of the Anandan quantum phase \cite{anan,anan2}, the He-McKellar-Wilkens effect \cite{hmw} and the scalar Aharonov-Bohm effect for neutral particles with a permanent electric dipole moment  \cite{zei,anan,anan2} can be obtained by defining different scenarios of the Lorentz symmetry breaking induced by the parity-odd sector of the tensor $\left(K_{F}\right)_{\mu\nu\alpha\beta}$.

\section{conclusions}

We have discussed new possible scenarios of studying the violation of the Lorentz symmetry based on the appearance of the relativistic geometric quantum phases in the wave function of a Dirac neutral particle induced by the parity-even and the parity-odd sectors of the tensor $\left(K_{F}\right)_{\mu\nu\alpha\beta}$ that governs the Lorentz symmetry violation in the CPT-even gauge sector of the Standard Model Extension. We have shown that it is possible to obtain a relativistic analogue of the Anandan quantum phase \cite{anan,anan2} induced by the parity-even sector of the tensor $\left(K_{F}\right)_{\mu\nu\alpha\beta}$, whose geometric phase is a non-Abelian geometric phase and a non-dispersive quantum phase. As a particular case, we have shown that a relativistic analogue of the Aharonov-Casher effect \cite{ac} can be achieved in a Lorentz symmetry violation scenario defined by a radial electric field produced by a linear distribution of electric charges and by considering the matrix $\left(\kappa_{HB}\right)_{ij}$ to be a diagonal matrix. In this case, we have also shown that bounds for the Lorentz symmetry breaking parameters can be calculated as given in Eq. (\ref{3.6}). Another particular case has been achieved in a Lorentz symmetry violation scenario defined by a uniform magnetic field along the $z$-axis and by considering the matrix $\left(\kappa_{HB}\right)_{ij}$ to be a diagonal matrix, which corresponds to the relativistic analogue of the scalar Aharonov-Bohm effect for a neutral particle with a permanent magnetic dipole moment \cite{zei,anan,anan2}. Again, we have shown that bounds for the Lorentz symmetry breaking parameters can be calculated as given in Eq. (\ref{3.9}).

Furthermore, we have shown that it is possible to obtain a relativistic analogue of the Anandan quantum phase \cite{anan,anan2} induced by the parity-odd sector of the tensor $\left(K_{F}\right)_{\mu\nu\alpha\beta}$. Again, we have obtained a non-Abelian and non-dispersive relativistic geometric phase. We have also discussed particular cases of the relativistic Anandan quantum phase based on two different scenarios of the Lorentz symmetry breaking. Based on a Lorentz symmetry violation scenario defined by a radial magnetic field produced by a linear distribution of magnetic charges and by considering the matrix $\left(\kappa_{DB}\right)_{ij}$ to be a diagonal matrix, we have shown that a relativistic analogue of the He-McKellar-Wilkens effect \cite{hmw} can be achieved. Finally, we have considered a Lorentz symmetry violation scenario defined by a uniform electric field along the $z$-axis and the matrix $\left(\kappa_{HE}\right)_{ij}$ to be a diagonal matrix. This last case corresponds to the relativistic analogue of the scalar Aharonov-Bohm effect for a neutral particle with a permanent electric dipole moment \cite{zei,anan,anan2}.

\acknowledgments

The authors would like to thank CNPq (Conselho Nacional de Desenvolvimento Cient\'ifico e Tecnol\'ogico - Brazil) for financial support.

\end{document}